\RequirePackage{ifpdf}
\ifpdf % We are running pdfTeX in pdf mode
\documentclass[pdftex]{sigma}
\else
\documentclass{sigma}
\fi

\newcommand{\hook}{\raisebox{-0.35ex}{\makebox[0.6em][r]
{\scriptsize $-$}}\hspace{-0.15em}\raisebox{0.25ex}
{\makebox[0.4em][l]{\tiny $|$}}}

\begin{document}

\allowdisplaybreaks

\renewcommand{\thefootnote}{$\star$}

\renewcommand{\PaperNumber}{037}

\FirstPageHeading

\ShortArticleName{Covariant Approach of the Dynamics of Particles}

\ArticleName{Covariant Approach of the Dynamics of Particles\\ in
External Gauge Fields, Killing Tensors\\ and Quantum
Gravitational Anomalies\footnote{This
paper is a contribution to the Proceedings of the Conference ``Symmetries and Integrability of Dif\/ference Equations (SIDE-9)'' (June 14--18, 2010, Varna, Bulgaria). The full collection is available at \href{http://www.emis.de/journals/SIGMA/SIDE-9.html}{http://www.emis.de/journals/SIGMA/SIDE-9.html}}}

\Author{Mihai VISINESCU}

\AuthorNameForHeading{M.~Visinescu}

\Address{Department of Theoretical Physics,
National Institute for Physics and Nuclear Engineering,\\
P.O. Box M.G.-6, Magurele, Bucharest, Romania}
\Email{\href{mailto:mvisin@theory.nipne.ro}{mvisin@theory.nipne.ro}}
\URLaddress{\url{http://www.theory.nipne.ro/~mvisin/}}

\ArticleDates{Received February 02, 2011, in f\/inal form March 28, 2011;  Published online April 05, 2011}

\Abstract{We give an overview of the f\/irst integrals of motion of particles in
the presence of external gauge f\/ields in a covariant Hamiltonian
approach. The special role of St\"ackel--Killing and Killing--Yano tensors
is pointed out. Some nontrivial examples involving Runge--Lenz type
conserved quantities are explicitly worked out. A condition of
the electromagnetic f\/ield to maintain the hidden symmetry of the system
is stated. A concrete realization of this condition is given by the
Killing--Maxwell system and exemplif\/ied with the Kerr metric. Quantum
symmetry operators for the Klein--Gordon and Dirac equations are
constructed from Killing tensors. The transfer of the classical
conserved quantities to the quantum mechanical level is analyzed in
connection with quantum anomalies.}


\Classification{81T20; 81T50}

\section{Introduction}

Symmetries comprise the most fundamental laws of nature, allowing to
f\/ind conserved quantities and exact solutions of the equations of
motion. The existence of a suf\/f\/icient number of conserved quantities
facilitates the investigation of a given dynamical system. Complete
integrability of the Hamiltonian  systems is closely related with the
separation of variables of the Hamilton--Jacobi equations \cite{SB}.

The customary conserved quantities originate from geometrical
symmetries of the conf\/i\-gu\-ra\-tion space of the system. These symmetries
correspond to Killing (K) vectors representing the isometries of the
spacetimes. Beside these symmetries, there exist hidden symmetries
ge\-ne\-ra\-ted by higher rank St\"ackel--Killing (SK) tensors. The
corresponding conserved quantities are quadratic, or, more general,
polynomial in momenta connected with symmetries of the complete
phase-space.

Another natural generalization of the Killing vectors is
represented by the antisymmetric Killing--Yano (KY) tensors \cite{KY}.
The `square' of a KY tensor is a SK tensor, but the opposite is not
generally true.
The symmetries generated by KY tensors are fundamental being involved
in many conserved quantities of the quantum systems. For example KY
tensors appear in the description of spinning particle systems~\cite{GRH}, construction of dif\/ferential operators which commute
with the Dirac operator~\cite{CML}, generation of new exotic
supersymmetries~\cite{MC}, and so on.

The conformal extension of the Killing vectors and SK, KY tensors is given by
conformal Killing (CK) vectors, conformal St\"ackel--Killing (CSK) and
conformal Killing--Yano (CKY) tensors respectively. These geometrical
objects are involved in the dynamics of the massless particles and are
connected with the f\/irst integrals of the null geodesics. In recent
times the properties of the CKY tensors stimulated much work in
generating background metrics with black-holes solutions in
higher-dimensional spacetimes (see, e.g.~\cite{VPF}) or interesting
geometrical structures~\cite{IVV}.

In the study of the dynamics of particles in external gauge f\/ields, the
covariant Hamiltonian formulation proposed by van Holten \cite{vH}
proves to be more adequate involving gauge covariant equations of
motion. In particular this approach permits the investigation of the
possibility for a~higher order symmetry to survive when the
electromagnetic interactions are taken into account. This possibility
is realized in the Killing--Maxwell (KM) system~\cite{BC} and an
explicit example is provided by the Kerr metric.

An unavoidable problem is the investigation to what extent the
classical conservation laws in a curved spacetime could be associated
with quantum symmetry operators. It is shown that the CK vectors and
CSK tensors do not in general produce quantum operators for the
Klein--Gordon equation. In connection with Dirac type operators
constructed from KY tensors we discuss the axial anomalies.

The plan of this review paper is as follows. In Section~\ref{section2} we present
the generalized Killing equations in a covariant framework including
external gauge f\/ields and scalar potentials. In the next section we
produce some examples of conserved quantities of Runge--Lenz type
involving external electromagnetic f\/ields. In Section~\ref{section4} we present
various generalizations of the Killing vectors.
In Section~\ref{section5} we analyze the possibility for a hidden
symmetry to survive when the electromagnetic interaction is taken into
account. We describe the KM system giving a concrete realization in the
Kerr spacetime. In Section~\ref{section6} we examine the quantum version of the
hidden symmetries and analyze the gravitational and axial anomalies
connected with quantum symmetry operators for Klein--Gordon and Dirac
equations. The last section is devoted to conclusions.

\section{Symmetries and conserved quantities}\label{section2}

The classical dynamics of a point charged particle subject to an
external electromagnetic f\/ield expressed (locally) in terms of
the potential $1$-form~$A_{i}$
\begin{gather}\label{FdA}
F=dA ,
\end{gather}
is derived from the Hamiltonian
\begin{gather}\label{H2}
H = \frac{1}{2} g^{ij} (p_i - A_i) (p_j - A_j) + V(x) .
\end{gather}
We also added an external scalar potential $V(x)$ for later
convenience and~$\mathbf{g}$ is the metric of a~(pseudo-)Riemmanian
$n$-dimensional  manifold~$\mathcal{M}$.

In terms of the canonical phase-space coordinates $(x^i, p_i)$ the
conserved quantities commute with the Hamiltonian in the sense of
Poisson brackets. The disadvantage of the traditional approach is that
the canonical momenta~$p_i$
and implicitly the Hamilton equations of motion are not manifestly
gauge covariant. This inconvenience can be removed using van Holten's
receipt~\cite{vH} by introducing the gauge invariant momenta:
\begin{gather*}%\label{Pi}
\Pi_i = p_i - A_i .
\end{gather*}
The Hamiltonian \eqref{H2} becomes
\begin{gather}\label{Hcov}
H = \frac{1}{2} g^{ij} \Pi_i \Pi_j  + V(x) ,
\end{gather}
and equations of motion are derived using the Poisson bracket
\begin{gather}\label{covPB}
\{P,Q\} = \frac{\partial P}{\partial x^i} \frac{\partial Q}{\partial
\Pi_i} -\frac{\partial P}{\partial \Pi_i} \frac{\partial Q}{\partial x^i}
+ q F_{ij}\frac{\partial P}{\partial \Pi_i}
\frac{\partial Q}{\partial \Pi_j} .
\end{gather}
We mention that in the modif\/ied Poisson bracket
the momenta $\Pi_i$ are not canonical.

A f\/irst integral of degree $p$ in the momenta $\Pi$ is of the form
\begin{gather}\label{cq}
K = K_0  + \sum^{p}_{k=1}\frac{1}{k!} K^{i_1 \dots i_k}_k
\Pi_{i_1} \cdots \Pi_{i_k},
\end{gather}
and it  has vanishing Poisson bracket \eqref{covPB} with the
Hamiltonian, $\{K,H\} = 0$, which implies
\begin{subequations}\label{constr}
\begin{gather}
K^i_1 V_{,i} = 0 , \label{1}\\
K_{0}^{~,i} +   F_{j}^{~i} K^j_1 = K^{ij}_2 V_{,j} , \label{0}\\
K^{(i_1 \dots i_l;i_{l+1})}_l + F_j^{~(i_{l+1}}
K^{i_1 \dots i_l) j}_{l+1}
= \frac{1}{(l+1)}  K^{i_1 \dots i_{l+1}j}_{l+2} V_{,j}
\qquad \mbox{for} \quad l= 1,\dots,p-2 ,  \label{l} \\
K^{(i_1 \dots i_{p-1};i_p)}_{p-1} + F_j^{~(i_p}
K^{i_1 \dots i_{p-1}) j}_p =0 ,\label{p-1}\\
K^{(i_1 \dots i_p;i_{p+1})}_p =0 . \label{p}
\end{gather}
\end{subequations}
Here a semicolon denotes the covariant dif\/ferentiation corresponding
to the Levi-Civita connection  and round brackets denote full
symmetrization over the indices enclosed.

The last equation~\eqref{p} is the def\/ining equation of a
SK tensor of rank~$p$. The SK tensors represent a
generalization of the Killing vectors and are responsible for the hidden
symmetries of the motions, connected with conserved quantities of the
form~\eqref{cq} polynomials in momenta. Indeed, using equation~\eqref{p}, for any geodesic $\gamma$ with tangent vector ${\dot x}^i
= p^i$
\begin{gather}\label{QK}
Q_K = K^{i_1 \dots i_k}_k p_{i_1} \cdots p_{i_k} ,
\end{gather}
is constant along $\gamma$.
The rest of the equations~\eqref{constr} mixes up the terms of $K$ with
the gauge f\/ield strength~$F_{ij}$ and derivatives of the potential~$V(x)$.
Several applications using van Holten's covariant framework \cite{vH}
are given in \cite{HN,JPN,MV1,IKI} and a few will be presented in the
next section.

\section{Explicit examples}\label{section3}

Let us illustrate these general considerations by some nontrivial
examples. In what follows the Coulomb potential in a $3$-dimensional
Euclidean space $\mathbb{E}^3$ will be the basis of the examples
superposing dif\/ferent types of electric and magnetic f\/ields.
The hidden symmetries which will be found involve SK tensors of rank
$2$ looking for constants of motion of the form
\begin{gather}\label{RLE}
K = K_0 + K^{i}_1 \Pi_i + \frac{1}{2} K^{ij}_2 \Pi_i \Pi_j .
\end{gather}

\subsection{Coulomb potential}

To put in a concrete
form, we consider the Hamiltonian for the motion of a point charge $q$
of mass $M$ in the Coulomb potential  produced by a charge $Q$
\[
H = \frac{M}{2} \dot{\mathbf{x}}^2 + q\frac{Q}{r} .
\]

We start with \eqref{p} for $p=2$ which is satisf\/ied by a SK tensor of
rank $2$. The most  general rank $2$ SK tensor on
$3$-dimensional Euclidean space involves some terms which do not
contribute nothing of interest for the Coulomb problem and
it proved that the following form of the
SK tensor is adequate \cite{Crampin}:
\begin{gather}\label{K2}
K^{ij}_2 = 2 \delta^{ij} \mathbf{n}\cdot \mathbf{x} -
\big(n^i x^j + n^j x^i\big) ,
\end{gather}
with $\mathbf{n}$ an arbitrary constant vector.

Corresponding to this SK tensor
the non-relativistic Coulomb problem admits the Runge--Lenz vector
constant of motion
\begin{gather}\label{RL}
\mathbf{K}_2 = \mathbf{p} \times \mathbf{L} +
MqQ\frac{\mathbf{x}}{r} ,
\end{gather}
where
\begin{gather}\label{am}
\mathbf{L} =\mathbf{x} \times \mathbf{p} ,
\end{gather}
is the angular momentum and $\mathbf{p} = M \dot{\mathbf{x}}$.

\subsection{Constant electric f\/ield}
The next more involved example consists of an electric charge $q$
moving in the Coulomb potential in the presence of a constant electric
f\/ield $\mathbf{E}$. The corresponding Hamiltonian is:
\[
H = \frac{1}{2 M} \mathbf{p}^2 + q\frac{Q}{r} -
q \mathbf{E}\cdot\mathbf{x}.
\]

Again it is adequate to take for the SK tensor of rank $2$ the
simple form  \eqref{K2} choosing $\mathbf{n}=\mathbf{E}$.
Using this form for $K^{ij}_2$ after a straightforward working out~\eqref{0}
\[
K_0 = \frac{MqQ}{r} \mathbf{E}\cdot \mathbf{x} -
\frac{Mq}{2} \mathbf{E}\cdot [\mathbf{x} \times
(\mathbf{x} \times \mathbf{E})] .
\]

Concerning equation \eqref{1}, it is automatically satisf\/ied by
a vector $\mathbf{K}_1$ of the form
\[
\mathbf{K}_1 = \mathbf{x} \times \mathbf{E} ,
\]
modulo an arbitrary constant factor. This vector
$\mathbf{K}_1$
contribute to a conserved quantity with a~term proportional to the
angular momentum $\mathbf{L}$ along the direction of the electric f\/ield
$\mathbf{E}$.

In conclusion, when a uniform constant electric f\/ield is present, the
Cou\-lomb system admits two constants of motion
$\mathbf{L}\cdot\mathbf{E}$
and $\mathbf{C}\cdot\mathbf{E}$ where $\mathbf{C}$ is a generalization
of the Runge--Lenz vector~\eqref{RL}:
\[
\mathbf{C} = \mathbf{K}_2 - \frac{Mq}{2}
\mathbf{x} \times ( \mathbf{x} \times \mathbf{E}) .
\]

\subsection{Spherically symmetric magnetic f\/ield}
Another conf\/iguration which admits a hidden symmetry is the
superposition of an external spherically symmetric magnetic f\/ield
\[
\mathbf{B} = f(r) \mathbf{x} ,
\]
over the Coulomb potential acting on a electric charge $q$.
This conf\/iguration is quite similar to the Dirac
charge-monopole system.

For $K^{ij}_2$ we use
again the form \eqref{K2} and $F_{ij}$  in this case is
\[
F_{ij} = \epsilon_{ijk} B^k = \epsilon_{ijk} x^k f(r) .
\]

The system of constraint \eqref{constr}  can be solely solved
only for a def\/inite form of the function $f(r)$
\[
f(r) = \frac {g}{r^{5/2}} ,
\]
with
$g$ a constant connected with the strength of the magnetic f\/ield.

With this special form of the function $f(r)$ we get
\[
K_0 = \left [ \frac{MqQ}{r} - \frac{2g^2 q^2}{r}\right ](\mathbf{n}\cdot
\mathbf{x}) ,
\]
and
\[
K^{i}_1 = - \frac{2gq}{r^{1/2}}(\mathbf{x}\times \mathbf{n})^i .
\]

Collecting the terms $K_0$, $K_1^{i}$, $K_2^{ij}$ the constant of motion
\eqref{RLE} becomes
\begin{gather}\label{ssmf}
K = \mathbf{n}\cdot \left (\mathbf{K}_2 + \frac{2gq}{r^{1/2}}\mathbf{L}
 - 2 g^2 q^2 \frac{\mathbf{x}}{r}\right ) ,
\end{gather}
with $\mathbf{n}$ an arbitrary constant unit vector and $\mathbf{K}$,
$\mathbf{L}$ given by~\eqref{RL},~\eqref{am} respectively.
The angular momentum~$\mathbf{L}$  is not separately
conserved, entering the constant of motion~\eqref{ssmf}.

\subsection{Magnetic f\/ield  along a f\/ixed direction}

Another example consists in a magnetic f\/ield directed along a f\/ixed
unit vector $\mathbf{n}$
\[
\mathbf{B} = B(\mathbf{x} \cdot \mathbf{n}) \mathbf{n} ,
\]
where, for the beginning, $B(\mathbf{x} \cdot \mathbf{n})$ is an
arbitrary function.

Again we are looking for a constant of motion of the form \eqref{RLE}
with the SK tensor of rank~$2$~\eqref{K2}.
Equations~\eqref{constr}  prove to be solvable only for a
particular form of the magnetic f\/ield
\[
\mathbf{B} = \frac{\alpha}{\sqrt{\alpha \mathbf{x} \cdot \mathbf{n} +
\beta }}\, \mathbf{n} ,
\]
with $\alpha$ and $\beta$ two arbitrary constants.

Consequently  we get for $K_0$ and $K_1^{i}$
\begin{gather*}
K_0 = \frac{MqQ}{r} (\mathbf{x}\cdot \mathbf{n}) +
\alpha q^2 (\mathbf{x} \times \mathbf{n})^2  ,\qquad
K_1^{i} = - 2 q \sqrt{\alpha \mathbf{x} \cdot \mathbf{n} + \beta }\,
(\mathbf{x} \times \mathbf{n})_i .
\end{gather*}

The f\/inal form of the conserved quantity in this case is:
\[
K= \mathbf{n} \cdot \left [ \mathbf{K}_2 + 2 q\sqrt{\alpha \mathbf{x}
\cdot \mathbf{n} + \beta }\, \mathbf{L} \right ] +
\alpha q^2 (\mathbf{x} \times \mathbf{n})^2  .
\]

As in the previous example the angular momentum $\mathbf{L}$ is
forming part of this constant of motion~$K$.

\subsection{Taub-NUT space and its generalizations}

The four-dimensional Euclidean Taub-NUT geometry is involved in many
modern studies in physics (see e.g.~\cite{VV} and reference therein).
The Taub-NUT metric is Ricci-f\/lat self dual on~$\mathbb{R}^4$ and gives
an example of non-trivial gravitational instanton. A Kaluza--Klein
monopole was described by embedding the Taub-NUT gravitational
instanton into f\/ive-dimensional Kaluza--Klein theory. Since the
classical equations of motion contain the Dirac monopole, a Coulomb
potential and a velocity-square dependent term, the Taub-NUT system
represents a non-trivial generalization of the Coulomb/Kepler system.

More interestingly, the Kaluza--Klein monopole in classical and quantum
mechanics possesses conserved quantities that are analog of the angular
momentum and the Runge--Lenz vector of the Kepler problem. Also various
generalizations of the Kaluza--Klein monopole system are
superintegrable, multiseparable~\cite{IM}.

Let us consider the radially symmetric generalized Taub-NUT metric
\cite{IK}:
\begin{gather}\label{gTNUT}
ds^2 = f(r) \delta_{ij} dx^i dx^j + h(r)\big(dx^4  + A_k dx^k\big)^2 ,\qquad
i,j,k =1,2,3 ,
\end{gather}
where $r$ is the radial coordinate on $\mathbb{R}^4 - \{0\}$ and $A_k$
is the gauge potential of a Dirac monopole.

For the geodesic motion on the $4$-manifold endowed with the metric
\eqref{gTNUT} the canonical momenta conjugate to the coordinates
$(x^i, x^4)$ are
\begin{gather*}
p_j = f(r)\delta_{ij} \frac{d x^i}{d t} + h(r)\left (\frac{d x^4}{d t} +
A_k\frac{d x^k}{d t} \right ) A_j ,\qquad
p_4 = h(r)\left (\frac{d x^4}{d t} + A_k\frac{d x^k}{d t}\right ) = q .
\end{gather*}

Let us remark that the
momentum associated with the cyclic variable $x^4$ is conserved and
interpreted as {\it relative electric charge}. Geodesic motion on the
$4$-manifold projects onto the curved $3$-manifold with metric $g_{ij}
= f(r) \delta_{ij}$ augmented with a potential \cite{IK,JPN}. In the
Hamiltonian~\eqref{Hcov} the potential is
\begin{gather}\label{VTNUT}
V(r) =  \frac{q^2}{2 h(r)} ,
\end{gather}
and accordingly the conserved energy is
\[
E = \frac{\mathbf{\Pi}^2}{2 f(r)} + \frac{q^2}{2 h(r)} .
\]

Now the search of conserved quantities of motion in the
$3$-dimensional curved space in the presence of the potential~\eqref{VTNUT} proceeds as in the previous examples. The zero-order
consistency condition~\eqref{1} is satisf\/ied for an arbitrary radial
potential and entails the conserved angular momentum which involves a
typical monopole  term
\begin{gather}\label{JTNUT}
\mathbf{J} = \mathbf{x} \times \mathbf{\Pi} + q \frac{\mathbf{x}}{r} .
\end{gather}

Next we search for a Runge--Lenz type vector and for this purpose we
start again with the SK tensor of the form \eqref{K2}. The set of
equations \eqref{constr} could be solved in some favorable
circumstances. First of all, in the original Taub-NUT space
\[
f(r) = \frac{1}{h(r)} = 1 + \frac{4 m}{r} ,
\]
where $m$ is a real parameter, the Runge--Lenz type vector is
\begin{gather}\label{RLTNUT}
\mathbf{K}_2 = \mathbf{\Pi} \times \mathbf{J} - 4 m \big(E - q^2\big)
\frac{\mathbf{x}}{r} .
\end{gather}

Another notable case is represented by the Iwai--Katayama
generalizations of the Taub-NUT metric \cite{IK}
\[
f(r) = \frac{a+b r}{r}  ,\qquad  h(r) = \frac{a r + b r^2}{1+ d r + c r^2} ,
\]
with $a, b, c, d \in \mathbb{R}$. The corresponding Runge--Lenz type
vector is
\begin{gather}\label{RLgTNUT}
\mathbf{K}_2 = \mathbf{\Pi} \times \mathbf{J} - \left(a E -\frac{d}{2} q^2\right)
\frac{\mathbf{x}}{r} .
\end{gather}

In both cases, due to the simultaneous existence of the conserved
angular momentum \eqref{JTNUT} and the conserved Runge--Lenz vectors
\eqref{RLTNUT} and respectively \eqref{RLgTNUT} the motions of the
particles are conf\/ined to conic sections.

\section{Generalizations of the Killing vectors}\label{section4}

A vector f\/ield $X$ on $\mathcal{M}$
is said to be a Killing vector f\/ield if the Lie derivative with
respect to~$X$ of the metric $\mathbf{g}$ vanishes:
\[
L_Xg=0 .
\]

Killing vector  f\/ields can be generalized to CK vector
f\/ields \cite{KY}, i.e.\ vector f\/ields with a f\/low preserving a given
conformal class of metrics. Furthermore a natural generalization of
CK vector f\/ields is given by the CKY tensors
\cite{KSW}. A CKY tensor of
rank $p$ on a (pseudo-)Riemmanian manifold $(\mathcal{M},\mathbf{g})$
is a $p$-form $Y (p\leq n)$ which satisf\/ies:
\begin{gather}\label{CKY}
\nabla_X Y = \frac{1}{p+1} X\hook dY - \frac{1}{n-p+1} X^\flat \wedge
d^* Y ,
\end{gather}
for any vector f\/ield $X$ on $\mathcal{M}$, where $\nabla$ is the
Levi-Civita connection of $\mathbf{g}$, $X^\flat$ is the 1-form
dual to the vector f\/ield $X$ with respect to the metric,
$\hook$ is the operator dual to the wedge product and $d^*$ is
the adjoint of the exterior derivative $d$.
Let us
recall that the Hodge dual maps the space of $p$-forms into the space
of $(n-p)$-forms. The square of $*$ on a $p$-form $Y$ is either~$+1$ or~$-1$ depending on $n$, $p$ and the signature of the metric~\cite{JK,MC}
\[
**Y = \epsilon_p Y  ,\qquad *^{-1} Y = \epsilon_p * Y ,
\]
with the number $\epsilon_p$
\[
\epsilon_p = (-1)^p *^{-1}\frac{\det g}{\vert \det g \vert} .
\]

With this convention, the exterior co-derivative can be written in terms
of $d$ and the Hodge star:
\[
d^* Y = (-1)^p *^{-1} d * Y .
\]

If $Y$ is co-closed
in \eqref{CKY}, then we obtain the def\/inition of a KY tensor~\cite{KY}
\begin{gather}\label{KY}
\nabla_X Y = \frac{1}{p+1} X\hook dY .
\end{gather}
This def\/inition is equivalent with the property that
$\nabla_j Y_{i_1 \dots i_p}$ is totally antisymmetric or, in
components,
\[
Y_{i_1 \dots i_{p-1}(i_p;j)} = 0 .
\]

The connection with the symmetry properties of the geodesic
motion is the observation that along every geodesic $\gamma$ in
$\mathcal{M}$, $Y_{i_1 \dots i_{p-1}j} \dot{x}^j$
is parallel.

There is also a conformal generalization of the SK tensors, namely a
symmetric tensor $K_{i_1 \dots i_p}= K_{(i_1 \dots i_p)}$ is
called a conformal Killing (CSK) tensor if it obeys the equation
\begin{gather}\label{CSK}
K_{(i_1 \dots i_p;j)} = g_{j(i_1}\tilde{K}_{i_2
\dots i_p)} ,
\end{gather}
where the tensor $\tilde{K}$ is determined by tracing the both sides of
equation~\eqref{CSK}. Let us note that in the case of CSK tensors, the
quantity~\eqref{QK} is constant only for null geodesics~$\gamma$.

These  generalizations of the Killing vectors could be
related. Let $Y_{i_1 \dots i_p}$ be a (C)KY tensor, then the
symmetric tensor f\/ield
\begin{gather}\label{KYY}
K_{ij} = Y_{i i_2 \dots i_p}Y_{j}^{~i_2 \dots i_p} ,
\end{gather}
is a (C)SK tensor and it sometimes refers to this (C)SK tensor as the
associated tensor with $Y_{i_1 \dots i_p}$. That is the case of
the Kerr metric \cite{RF,RP} or the Euclidean Taub-NUT space
\cite{GR,VV}. However, the converse statement is not true in general:
not all SK tensors of rank $2$ are associated with a KY tensor.
To wit in the Taub-NUT geometry there are known to exist four KY tensors of
which three are covariantly constant. The components of the Runge--Lenz
vector \eqref{RLTNUT} are SK tensors which are associated with the KY
tensors of the Taub-NUT space. On the other hand, in the case of the
Runge--Lenz vector \eqref{RLgTNUT} its components  are also SK tensors
but not associated with KY tensors since the generalized Taub-NUT space
\cite{IK} does not admit KY tensors \cite{MV3}.

Drawing a parallel between def\/initions \eqref{CKY} and \eqref{KY} we
remark that all KY tensors are co-closed but not necessarily closed.
From this point of view CKY tensors represent a generalization  more
symmetric in the pair of notions. CKY equation \eqref{CKY} is invariant
under Hodge duality that if a $p$-form $Y$ satisf\/ies it, then so does
the $(n-p)$-form $*Y$. Moreover the dual of a CKY tensor is a KY tensor
if and only if it is closed.

Let us assume that a CKY tensor of rank $p = 2$ is closed ($d Y = 0$)
and non-degenerate called a {\it principal CKY tensor}. The principal
CKY tensor obeys the following equation \cite{JJ,Frolov}
\[
\nabla_X Y = X^\flat \wedge \xi^\flat ,
\]
where $X$ is an arbitrary vector f\/ield and
\begin{gather}\label{pKV}
\xi_i = \frac{1}{n-1} \nabla_j Y^j_{~i}  .
\end{gather}
Starting with a principal CKY
tensor one can construct a tower of CKY tensors formed from external
powers  $Y^{\wedge k}$ which again are closed CKY tensors. Taking the
Hodge dual of these tensors one obtains a set of KY tensors. On the
other hand, the vector $\xi_i$ \eqref{pKV} obeys the following equation
\[
\xi_{(i;j)} = -\frac{1}{n-2} R_{l(i} Y_{j)}^{~~l} .
\]
It is obvious that in a Ricci f\/lat space ($R_{ij} = 0$) or in an
Einstein space ($R_{ij} \sim g_{ij}$), $\xi_{i}$ is a~Killing
vector and we shall refer to it as the {\it primary Killing vector}.

%%%%%%%%%%%%%%%%%%%%%%%%%%%%%%%%%%%%%%%%%%%%

\section[Killing-Maxwell system]{Killing--Maxwell system}\label{section5}

In this section we shall analyze what is the condition which must be
imposed on the gauge f\/ields to preserve the hidden symmetry of the
system. Examining the set of coupled equations~\eqref{constr}, we
observe that the vanishing of the terms
$F_j^{~(i_l} K^{i_1 \dots i_{l-1}) j} $ guarantees that the gauge
f\/ields do not af\/fect the symmetries of the system.
To make things more
specif\/ic, let us assume that the system admits a hidden symmetry
encapsulated in a SK tensor of rank~$2$, $K_{ij}$,
associated with a~KY tensor $Y_{ij}$ according to~\eqref{KYY}. The
suf\/f\/icient condition of the electromagnetic f\/ield to preserve the
hidden symmetry is \cite{MV1}
\begin{gather}\label{cond}
F_{k[i} Y_{j]}^{~k} = 0 ,
\end{gather}
where the indices in square bracket are to be antisymmetrized.

It is worth mentioning that this condition appear in many other
contexts. For example, using conformal Killing spinors~\cite{HPSW} this
constraint was resorted to maintain the constant of motion along a null
geodesic. Also in the case of the motion of pseudo-classical spinning
point particles, relation \eqref{cond} assures the preservation of the
non-generic supersymmetry associated with a KY tensor~\cite{MT}.
In the case of spinor f\/ields on curved space-times this condition is
necessary in the construction of Dirac-type operators that commute with
the standard Dirac operator~\cite{McLS}.

The KM system described by Carter \cite{BC} represent an interesting
realization of the condition~\eqref{cond}. Let us consider the source
equation of an electromagnetic f\/ield~$F_{ij}$ in $4$-dimensions
\begin{gather}\label{Max}
F^{ij}_{~~;j} = 4\pi j^i,
\end{gather}
and assume that the current $j^i$ is a primary Killing vector. Drawing a
parallel between equations~\eqref{pKV} and \eqref{Max} we conclude
that the electromagnetic f\/ield in the KM system is a CKY tensor which,
in addition, is a closed $2$-form, as in \eqref{FdA}. Therefore its
Hodge dual
\begin{gather}\label{YHF}
Y_{ij} = * F_{ij} ,
\end{gather}
is a KY tensor which generates a hidden symmetry \eqref{KYY} associated
with it. It is quite simple to verify that $F_{ij} Y_{k}^{~j} \sim
F_{ij} *F_{k}^{~j}$ is a symmetric matrix ( in fact proportional with
the unit matrix) making obvious that the constraint \eqref{cond} is
fulf\/illed.

We complete the description of the KM system observing that the
$2$-form $Y$ \eqref{YHF} can be written, at least locally, as
\[
Y = * d A ,
\]
the form $A$ being usually called a {\it KY potential}.

\subsection{The Kerr metric}

To exemplify the results presented previously, let us consider the Kerr
solution to the vacuum Einstein equations which in the Boyer--Lindquist
coordinates $(t, r, \theta, \phi)$ has the form
\[
g= - \frac{\Delta}{\rho^2}\big( d t - a \sin^2 \theta d \phi\big)^2 +
\frac{\sin^2 \theta}{\rho^2} \big[\big(r^2 + a^2\big) d \phi - a d t\big]^2 +
\frac{\rho^2}{\Delta} d r^2 + \rho^2  d \theta^2 ,
\]
where
\begin{gather*}
\Delta = r^2 + a^2 - 2 m r , \qquad
\rho^2 = r^2 + a^2 \cos^2 \theta .
\end{gather*}
This metric describes a rotating black hole of mass $m$ and angular
momentum $J=a m$.

As was found by Carter \cite{BC2}, the Kerr space admits the SK tensor
\begin{gather*}
K_{ij} dx^i dx^j =  - \frac{\rho^2 a^2 \cos^2 \theta}{\Delta} d r^2 +
\frac{\Delta a^2 \cos^2 \theta}{\rho^2}
\big( d t - a \sin^2 \theta d \phi\big)^2 \\
\phantom{K_{ij} dx^i dx^j =}{}  +\frac{r^2 \sin^2 \theta}{\rho^2} \big[{-} a d t +\big(r^2 + a^2\big) d \phi\big]^2 +
\rho^2 r^2 d \theta^2 ,
\end{gather*}
in addition to the metric tensor $g_{ij}$. This tensor is associated
with the KY tensor \cite{GRH,JL}
\[
Y = r \sin \theta d \theta \wedge \big[{-} a d t +\big(r^2 + a^2\big) d \phi\big]
+ a \cos \theta dr \wedge \big( d t - a \sin^2 \theta d \phi\big) ,
\]
and the  dual tensor
\[
*Y = a \sin \theta \cos \theta d \theta \wedge
\big[{-} a d t +\big(r^2 + a^2\big) d \phi\big]
+ r dr \wedge \big({-} d t + a \sin^2 \theta d \phi\big) ,
\]
is a CKY tensor.

The existence of this CKY tensor represents a realization of the KM
system and we can verify that the KM four-potential one-form is
\[
A = \frac{1}{2} \big( a^2 \cos^2 \theta - r^2\big) d t + \frac{1}{2} a \big( r^2 +
a^2\big) \sin^2 \theta d \phi .
\]
Finally, the current is to be identif\/ied with the primary Killing
vector \eqref{pKV}
\[
Y^{l}\partial_l := Y^{kl}_{~~;k} \partial_l = 3 \partial_t .
\]
%%%%%%%%%%%%%%%%%%%%%%%%%%%%%%%%%%%%%%%%%%%%

\section{Quantum anomalies}\label{section6}

In this section we shall establish a quantum version of the hidden
symmetries generated by Killing tensors. We have to be aware of the fact that
the classical conserved quantities may not be preserved when we pass
to the quantum mechanical level~-- that is {\it anomalies} may occur.
For the beginning we shall investigate the quantum symmetry operators
for the Klein--Gordon equation. Finally we def\/ine conserved operators in
the case of Dirac equation and analyze the axial anomalies for the
(generalized) Taub-NUT metric.

\subsection{Gravitational anomalies}

Let us now consider the necessary conditions for the existence of
constants of motion in the f\/irst-quantized system. We start with the
classical conserved quantities and take into account that in the
quantized system the momentum operator is given by the covariant
dif\/ferential operator~$\nabla$ on the manifold $\mathcal{M}$. The
corresponding Hamilton operator for a free scalar particle is given by
the covariant Laplacian acting on scalars:
\begin{gather}\label{KG}
{\cal H} = \square = \nabla_i g^{ij} \nabla_j = \nabla_i
\nabla^i ,
\end{gather}
and for a (C)K vector we def\/ine the quantum symmetry operator
\[
{\cal Q}_{V} = K^i \nabla_i .
\]

To consider the analogous condition for the existence of constants of
motion in the f\/irst-quantized system, we must evaluate the commutator
$[\square,{\cal Q}_{V}] \Phi$ acting on some function
$\Phi \in {\cal C}^\infty (\mathcal{M})$, solutions of the
Klein--Gordon equation $\square \Phi = 0$. The evaluation of this
commutator gives:
\[
[{\cal H},{\cal Q}_{V}] = \frac{2-n}{n} K_k^{~;ki}
\nabla_i + \frac{2}{n} K^k_{~;k} \square .
\]

In the case of ordinary Killing vectors, the r.h.s.\ of this commutator
vanishes on the strength of Killing equation, but for CK vectors the
situation is dif\/ferent. Since the term $K_k^{~;ki} \nabla_i$ survives
for CK vectors, in general the system is af\/fected by quantum anomalies.

For (C)SK tensors the situation is more intricate. The quantum analogue
of classical conserved quantity for a (C)SK tensor of rank $2$ is
\[
{\cal Q}_{T} = \nabla_i K^{ij} \nabla_j .
\]
On a general manifold this operator does not commute with the
Klein--Gordon operator \eqref{KG} \cite{IVV, MV2}:
\begin{gather}
[\square, {\cal Q}_{T}] = 2 \big( \nabla^{(k} K^{ij)}\big)
\nabla_{k} \nabla_{i} \nabla_{j}
+ 3 \nabla_{m} \big(\nabla^{(k} K^{mj)}\big)
\nabla_{j}\nabla_{k}  \nonumber\\
{+} \left \{  -\frac{4}{3} \nabla_{k}
\big( R_{m}^{~[k} K^{j]m}\big) +
\nabla_{k}\left ( \frac{1}{2}g_{ml}
(\nabla^{k}\nabla^{(m} K^{lj)} -
\nabla^{j}\nabla^{(m} K^{kl)} ) +
\nabla_{i}\nabla^{(k} K^{ij)}\right )\right \}
\nabla_{j} . \label{comuttotal}
\end{gather}
We mention that the last term is missing in the corresponding equation
in \cite{BC1}.

Some simplif\/ications occur for SK tensor since all the symmetrized
derivatives vanish and we end up with a simpler result \cite{BC1}
\[
[\square, {\cal Q}_{T}] = - \frac{4}{3} \nabla_{i}
\big(R^{~~[i}_{l} K^{j]l}\big) \nabla_{j} .
\]
There are a few notable situation for which the quantum system is free
of anomalies. Of course if the space is Ricci f\/lat or Einstein,
i.e.\ $R_{ij} \sim g_{ij}$ the r.h.s.\ of the commutator vanishes.
A more interesting and quite unexpected
case is represented by SK tensors associated to
KY tensors of rank $2$ as in~\eqref{KYY} \cite{BC1} a situation which
occurs for some spaces~\cite{RF,RP,GR,VV}.

In the case of CSK tensors, practically all terms in~\eqref{comuttotal}
survive. Taking into account that the terms are arranged into groups
with three, two and just one derivatives  it is
impossible to have compensations between them.

In conclusion CK vectors and CSK tensors do not in general produce
symmetry operators for the Klein--Gordon equation, the quantum system
being af\/fected by quantum anomalies.

\subsection{Dirac symmetry operators}

We shall assume that it is possible to def\/ine a Dirac spinor structure
on the base manifold~$\mathcal{M}$.
Spinor f\/ields carry irreducible representation of the Clif\/ford algebra
and in what follows we shall identify its elements with dif\/ferential
forms. Adopting the convention for the Clif\/ford algebra
\[
e^a e^b + e^b e^a = 2 g^{ab} ,
\]
the standard Dirac operator $D$ is written in terms of the spinor
connection as
\[
D = e^a \nabla_{X_a} .
\]

A key property of CKY tensors is that one may construct from them
symmetry operators for the massless Dirac equation \cite{Benn,CKK}.
Let us construct an operator acting on spinors:
\[
L_Y = e^a Y \nabla_{X_a} + \frac{p}{p+1} dY - \frac{n-p}{n-p+1} d^* Y .
\]
In order to construct a Dirac type operator, it proves to be convenient
to add to $L_Y$ a term proportional to the Dirac operator:
\[
D_Y = L_Y - (-1)^p Y D .
\]
This operator $D_Y$ is said to be $R$-(anti)commuting as the graded
commutator with the Dirac operator
\[
\{ D , D_Y\}_p = R D ,
\]
where the graded commutator is \cite{Benn,CKK}
\[
\{ D , D_Y\}_p = D D_Y + (-1)^p D_Y D .
\]
For the symmetry operator $D_Y$ the explicit form of the $R$ operator
is
\[
R = \frac{2 (-1)^p}{n-p+1} d^* Y D .
\]
Let us remark that the operator R vanishes when $Y$ is a KY tensor,
otherwise the  Dirac type operator $D_Y$ gives an {\it on-shell}
$(D\Psi=0$) symmetry operator for Dirac operator. The four dimensional
case was considered in \cite{KL}.

\subsection{Axial anomalies}

Having in mind that the KY tensors prevent the appearance of
gravitational anomalies for the scalar f\/ield, it is natural to
investigate whether they play a role also to {\it axial anomalies}.
Various authors have discussed the electromagnetic and gravitational
anomalies in the divergence of the axial-current which is closely
related to the existence of zero-eigenvalues of the Dirac operator.
The index of the Dirac operator is useful as a tool to investigate
topological properties of the space as well as in computing anomalies
in quantum f\/ield theories.

In what follows we shall consider even-dimensional spaces in which one
can def\/ine the index of a Dirac operator as the dif\/ference in the
number of linearly independent zero modes with eigenvalues $+1$ and
$-1$ under $\gamma_5$:
\[
index(D) = n^0_+ - n^0_- .
\]

The commutation relation between the standard and non-standard Dirac
operators generated by KY tensor
\[
[D_Y , D] = 0 ,
\]
leads to a remarkable result for the index of non-standard Dirac
operator:
\begin{theorem}
\[
{\rm index}(D_Y) = {\rm index}(D) .
\]
\end{theorem}
\begin{proof}
For a sketch of the proof see~\cite{HWP}.
\end{proof}

Since we analyzed various properties of the Taub-NUT metric and its
generalizations, in what follows we shall also consider the problem
of axial anomalies for this space. The remarkable result is that the
Taub-NUT metric makes no contribution to the axial anomaly:
\begin{theorem}
The Dirac operator associated with the standard Taub-NUT metric on
$\mathbb{R}^4$ does not admit any $L^2$ zero modes.
\end{theorem}

\begin{proof}
We sketch the proof \cite{MV} observing that the scalar curvature
$\kappa$ of the standard Taub-NUT metric vanishes. By the Lichnerowicz
formula
\[
D^2 = \nabla^* \nabla + \frac{\kappa}{4} = \nabla^* \nabla .
\]
Let $\Psi \in L^2 $ be a solution of $D$, hence $\nabla \Psi = 0$.
Since a parallel spinor has a constant pointwise norm, it cannot be in
$L^2$ unless it is $0$, because the volume of $\mathbb{R}^4$ with
respect to the Taub-NUT metric is inf\/inite. Therefore $\Psi = 0$.
\end{proof}

We turn now to the generalized Taub-NUT metric. It has proved in
\cite{MV} that on the whole generalized Taub-NUT space, although the
Dirac operator is not Fredholm, it only has a f\/inite number of null
states. In~\cite{MM} there were found suf\/f\/icient conditions for the
absence of harmonic~$L^2$ spinors on Iwai--Katayama~\cite{IK}
generalizations of the Taub-NUT space.

\begin{theorem}
There do not exist $L^2$ harmonic spinors on $\mathbb{R}^4$ for the
generalized Taub-NUT metrics. In particular, the $L^2$ index of the
Dirac operator vanishes.
\end{theorem}

\begin{proof}
We delegate the proof to \cite{MM}.
\end{proof}

%%%%%%%%%%%%%%%%%%%%%%%%%%%%%%%%%%%%%%%%%%%

\section{Concluding comments}\label{section7}

The (C)SK and (C)KY tensors are related to a multitude of dif\/ferent
topics such as classical integrability of systems together with their
quantization, supergravity, string theories, hidden symmetries in
higher dimensional black-holes spacetimes, etc.

To conclude let us discuss shortly some problems that deserve a further
attention. An obvious extension of the gauge covariant approach to
hidden symmetries is represented by the non-abelian dynamics using the
appropriate Poisson brackets \cite{vH,HN}. In Section~\ref{section3} we worked out
some examples in an Euclidean $3$-dimensional space and restricted to
SK tensors of rank~$2$. More elaborate examples working in a $N$-dimensional curved space
and involving SK tensors of higher ranks will be presented elsewhere~\cite{MVprep}.

Finally, let us mention that the extension of the (C)KY symmetry on
spaces with a skew-symmetric torsion is desirable and may provide new
insight into the theory of black holes \cite{HKWY}.

\subsection*{Acknowledgements}
Support through CNCSIS program IDEI-571/2008  is acknowledged.

\pdfbookmark[1]{References}{ref}
\LastPageEnding

\end{document}